\documentclass[a4paper,12pt]{article}

\usepackage{hyperref}
\usepackage{natbib}

\usepackage{amsmath}
\numberwithin{equation}{section}
\usepackage{amssymb}

\makeatletter
\renewcommand\@biblabel[1]{(#1)}
\makeatother

\begin{document}

\title{\bf Simulating leaky integrate and fire neuron with integers}

\author{A. K. Vidybida%
\thanks{Bogolyubov Institute for Theoretical Physics, Metrologichna str., 14-B, Kyiv 03680, Ukraine, vidybida@bitp.kiev.ua, http://www.bitp.kiev.ua/pers/vidybida}
}
\date{}

\maketitle

\begin{abstract}
The leaky integrate and fire (LIF) neuron represents standard neuronal model
used for numerical simulations. The leakage is implemented in the model
as exponential decay of trans-membrane voltage towards its resting value.
This makes inevitable the usage of machine floating point numbers in the
course of simulation.
It is known that machine floating point arithmetic
is subjected to small inaccuracies, which prevent 
from exact comparison of floating point quantities.
In particular, it is incorrect to decide whether two
separate in time states of a simulated system composed of LIF neurons
are exactly identical.
However, decision of this type is necessary, e.g.
to figure periodic dynamical regimes in a reverberating network.
Here we offer a simulation paradigm of a LIF neuron,
in which neuronal states are described by
whole numbers. Within this paradigm,
the LIF neuron behaves exactly the same way as does
the standard floating point simulated LIF, although
exact comparison of states becomes correctly defined.
\end{abstract}

\noindent
{\bf Keywords.} 
Leaky integrate and fire neuron; Floating point calculations; Simulations

\section{Introduction}
Leaky integrate and fire (LIF) neuron, \cite{Knight1972,Stein1967} 
is a universally-accepted ``standard'' neuronal model
in theoretical neuroscience. Usage of LIF as a model allowed 
obtaining numerous theoretical results, see e.g. 
\cite{Burkitt,Burkitta,Lansky1984,Tuckwell1988}.

When input impulses are absent,  the membrane potential of the LIF model
 decays exponentially.
In computer simulations, this makes inevitable the usage of 
machine floating point arithmetic.
However, 
the floating point calculations
can be considerably inaccurate in some cases, see e.g. \cite{Goldberg:1991:CSK:103162.103163,Hayes2003}. These cases do not so
frequently occur. Therefore, if one studies neuronal firing statistics,
when a simulation is repeated many times with slightly different
parameters, the errors due to floating point pitfalls can be negligibly
small. 

The problem arises if one needs to check if two neuronal
states obtained in the course of simulation are identical. 
This kind
of check is necessary, e.g. to figure periodic dynamical 
regimes of a reverberating network, see \cite{Vidybida2011} for an example. 
In order to check if a neural net is in
a periodic regime, one needs to check whether two states the net
passes through at two distinct moments of time are exactly the same.
If states of individual neurons the net is composed of are
described by floating point variables, then
this expects checking that
two floating point numbers are exactly equal --- the operation,
which is not permitted in programming due to small, 
but inevitable inaccuracy adhered to the machine floating point arithmetic. 
Instead, the option is to check whether
the distance between two numbers is less than a "machine
epsilon", \cite{Higham:2002:ASN}. This does not help to find periodic
regimes reliably. Indeed, in this comparison paradigm, two different states
may happen to be treated as identical and this may bring about a
fake periodic regime. Taking into account that rounding mode
depends on operating system and on machine architecture,
it will be difficult to treat in a consistent manner
sustained behavior of a dynamical system prone to instability
(as do neuronal nets) if simulation is made in floating point numbers.

This problem is avoided in \cite{Vidybida2011}, where the binding
neuron  model was used as neuronal model. The binding
neuron states are naturally described by integers, and machine arithmetic
of integers is perfectly accurate. This admits
a routine check whether two numbers are exactly equal, and finally,
whether two states of a neuronal net are identical, provided those states
are expressed in terms of whole numbers, exclusively.
It would be nice to have the same possibility when a LIF model
is used for simulations.

In traditional computer simulations of LIF model driven with input 
 stream of stimulating impulses, a dynamical system is simulated,
which has states described with machine floating point numbers. 
We denote this dynamical system as fpLIF.
The purpose of this paper is to offer an approximation,
which replaces the fpLIF with intLIF --- a dynamical system
whose states are expressed in terms of integers.
In what follows, we describe the approximation itself and check
its quality by running both fpLIF and intLIF with the same
stream of input impulses randomly distributed in time.
 It appears that it is possible to
choose the approximation parameters in such a way, that
dynamics of both fpLIF and intLIF are exactly the same, if
considered in terms of spiking moments.

\section{Methods}
\subsection{The fpLIF model}
The simplest LIF model is considered. Namely,
the neuronal state at moment $t$ is described as membrane voltage, 
$V(t)$. The resting state is defined as $V(t)=0$. Any input impulse
advances membrane potential by $h$, where $h>0$. Between two 
consecutive input impulses, $V(t)$ decays exponentially:
$$
V(t+s)=e^{-s/\tau}V(t)
$$
where $s>0$, $\tau$ --- is the membrane relaxation time.
Suppose that the threshold value for the LIF neuron is $V_0$.
The neuron fires a spike every time when 
$
V(t)\ge V_0,
$
and $V(t)$ is reset to zero after that. The set of possible values
of $V(t)$ is the following interval:
\begin{equation}\label{Interval}
V(t)\in[0;V_0[
\end{equation}
The above mentioned properties of the LIF model
can be routinely coded with $V(t)$ and $h$
declared as floating point quantities. In this case, possible
values of $V(t)$ will be those machine floating point numbers,
which fall into the interval (\ref{Interval}).
And this gives the fpLIF dynamical system.
\medskip

\subsection{The intLIF model}

\subsubsection{Pure decay dynamics}\label{Pdd}

In numerical simulation of a dynamical system, 
the time is advanced in discrete steps,
having duration of a small fixed\footnote{See Discussion for a remark about adjustable $dt$.} time-step, $dt$. 
This gives an approximation of the continuous time $t$ with
discrete moments:
\begin{equation}\label{moments}
\mathbb{T}=
\{0,dt, 2\,dt,3\,dt, \dots\}
\end{equation}
Due to this fact,
the membrane voltage, $V(t)$, also changes in a discrete manner from step to
step. As a result, in a single run of the pure decay dynamics (without
input stimulation) the $V(t)$ will pass through only some discrete values,
missing the intermediate ones. Those discrete values can be chosen as an
approximation of the continuous interval (\ref{Interval}). It is clear
that the set of the discrete values mentioned depends on the initial
value of $V$. To be concrete, let us chose the following approximating set:
\begin{equation}\label{bins1}
\mathbb{V}_\alpha=
\{\alpha\,V_0,\alpha^2\,V_0,\alpha^3\,V_0,\dots\},\text{ where }
\alpha=e^{-dt/\tau}
\end{equation}
This induces the following representation of the interval (\ref{Interval}):
\begin{equation}\label{intervals1}
[0;V_0[=\{0\}\cup \,[\alpha V_0;V_0[\,\cup\, [\alpha^2 V_0;\alpha V_0[\,
\cup\dots
\end{equation}
Any value $V(t)\in [0;V_0[$ falls for some $n$ into interval
$[\alpha^{n+1} V_0;\alpha^n V_0[$ and we chose its left end as an 
approximation for that $V(t)$. The error of this approximation
is less than
\begin{equation}\label{error1}
\Delta V_0=(1-\alpha)V_0
\end{equation}

Now, if neuron has membrane potential 
$V(t)=V\in [\alpha^{n+1} V_0;\alpha^n V_0[$ and one intends to describe
by values from $\mathbb{V}_\alpha$ its consequent dynamics due to
pure decay, the following should be done.
The $V(t)$ value should be replaced with $V_d=\alpha^{n+1} V_0$, 
and the subsequent decay dynamics simply propels
$V(t)$ through values $\alpha V_d,\alpha^2 V_d,\alpha^3 V_d,\dots$ from
$\mathbb{V}_\alpha$. The state of neuron at each time step can be labeled
with integer $n$ by the following way: 
The state $V(t)=\alpha^{n+1} V_0$ obtains
label $n$, where $n\ge 0$. Now, the decay dynamics can be expressed in terms of $n$. Namely,
if at some discrete moment of time $t\in\mathbb{T}$ a state is labeled as $n$,
then the state at the next moment, $t+dt$, is labeled as $n+1$.
Thus, having in mind the correspondence:
$
n\quad \leftrightarrow \quad \alpha^{n+1}V_0
$,
the pure decay dynamics becomes as simple as adding unity 
to the state label at each time step.

For computer simulation purpose, it should be noticed, that the set 
$\mathbb{V}_\alpha$ is finite.
If the state label $n$ is declared in a program as \verb-int-, then 
$\mathbb{V}_\alpha$ has
\verb-INT_MAX-+1  elements, where \verb-INT_MAX-
is the largest value of \verb-int--type variable which can be represented
in the operating system.
Thus, with state labels of type \verb-int-, instead of (\ref{bins1}),
one has to chose as $\mathbb{V}_\alpha$ the following set of 
\verb-INT_MAX-+1 elements:
$
\mathbb{V}_\alpha=
\{\alpha\,V_0,\alpha^2\,V_0,\dots,\alpha^{\verb-INT_MAX-}\,V_0,0\},
$
where the value 0 is added to describe the resting state, which is attained 
just after firing. Consequently, (\ref{intervals1}) should be replaced
with the following:
\begin{equation}\label{intervals1INT_MAX}
[0;V_0[=[\alpha V_0;V_0[\cup [\alpha^2 V_0;\alpha V_0[
\cup\dots\cup \,[0;\alpha^{\verb-INT_MAX-}\,V_0[
\end{equation}

In a 64-bits OS, \verb-INT_MAX- = 2147483647. This imposes a limit on
possible duration of the pure decay evolution represented with 
\verb-int- type labels.
If the time step $dt=0.01$ ms, then  \verb-INT_MAX-$\cdot dt\sim$ 
350 minutes. Thus, description of neuronal state by \verb-int- type label
fails, if a LIF neuron starts, e.g. with $V(t)$ close to $V_0$ and does not receive excitatory stimulation for
longer than 350 minutes of real time.

In real networks,  stimulus-free period of a neuron involved in 
a useful task cannot be so long. 
Indirectly, a very crude upper bound for possible duration of 
stimulus-free period can be
estimated from duration of suppression of activity observed in some brain
networks, \cite{Ossandon}. Such a suppression can last up to 1000 ms.
Based on this value, 
 let us expect that our neuron receives
excitatory input 
impulses\footnote{See the next section for exact treatment of 
input impulses.} 
of amplitude $h$ 
with mean rate higher than $r\sim 0.1$ Hz. Before the first input
impulse, neuron is in the state "empty" with $V=0$.
After the first input impulse the state label 
$n_1\sim \log_\alpha(h/V_0)$ and this
value is small if compared with \verb-INT_MAX-. For example, for
$dt=0.01$ ms, $\tau=20$ ms, $h/V_0=0.01$ one obtains $n_1\sim 10^4$.
The label $n_1$ gets increment 1 after each time step $dt$ until the
next input impulse comes. The mean waiting time for the next stimulation,
if expressed in the $dt$ units,
is $1/(r\cdot dt)=10^6$. This means that at the moment of next stimulation,
the state label $n_2\sim n_1+10^6 \ll 10^7.$ After the next input impulse,
the state label $n_3<n_1$. This suggests that states with labels $n>10^7$
will be observed
quite rare with no chances to attain $n$ value close to the \verb-INT_MAX-
provided a neuron receives at least a moderate stimulation.

\subsubsection{Input impulse}\label{Ii}
Suppose that input impulse at moment $t$ advances by $h$ the membrane voltage
$V(t)$. If $V(t)\in\mathbb{V}_\alpha$, then $V(t)+h\in\mathbb{V}_\alpha$
does not hold in most cases. One needs to approximate
$V(t)+h$ by a suitable value from $\mathbb{V}_\alpha$, as described in
n. \ref{Pdd},
and then to proceed with pure decay dynamics
expressed in integers, as described in n. \ref{Pdd}, until the next input
impulse.

Preliminary tests of this scheme 
where performed with LIF neuron stimulated with Poisson stream
of input impulses. The standard floating point LIF (fpLIF) simulation
and integer LIF (intLIF) simulation,
as described here,  were performed with the same input streams.
The threshold manner of triggering
gives chances that firing moments of both models will be the same,
if expressed in whole number of $dt$.
Computer simulations performed show that
the firing moments of both models are indeed
the same for some initial period
of time after which they become different.
This is because the approximation error,
when approximating LIF state (voltage) by a value from 
$\mathbb{V}_\alpha$, builds up too fast with each input impulse. 
 In order to decrease the error, one needs more precise approximation of
the continuum set $[0;V_0[$, than that given by the discrete set 
$\mathbb{V}_\alpha$. The required approximation can be achieved by introducing
second order bins into the set $\mathbb{V}_\alpha$.

\subsubsection{Second order bins}

Let us choose an integer $N>1$ and divide each bin from 
(\ref{intervals1INT_MAX})
to $N$ equal second order bins\footnote{We do not consider the last bin from 
(\ref{intervals1INT_MAX}), because we do not expect the state $V$ will ever fall into it, except of just after firing, when $V=0$, exactly.}. This gives representation of $n$-th 
first order bin:
$$
[\alpha^{n+1} V_0;\alpha^n V_0[=
\bigcup_{i=0}^{N-1}
[\alpha^{n+1} V_0 + i\cdot c_n;\,\alpha^{n+1} V_0 + (i+1)c_n[
$$
where $c_n$ denotes the size of second order bin within the $n$-th 
first order bin $[\alpha^{n+1} V_0;\alpha^n V_0[$:
\begin{equation}\label{cn}
c_n=(\alpha^n V_0 - \alpha^{n+1} V_0)/N
\end{equation}
Now, if 
$V(t)\in [\alpha^{n+1} V_0 + i\cdot c_n;\,\alpha^{n+1} V_0 + (i+1)c_n[$,
then we chose $V_d=\alpha^{n+1} V_0 + i\cdot c_n$ as its approximation.
This results in the new set $\mathbb{V}_{\alpha,N}$ of possible values
for $V(t)$:
$$
\mathbb{V}_{\alpha,N}=\{0\}\cup
\quad
\bigcup_{n=0}^{\verb-INT_MAX- \,-1}
\quad
\bigcup_{i=0}^{N-1}
\{\alpha^{n+1} V_0+i\,c_n\}
$$
Any point, except of 0, in the set $\mathbb{V}_{\alpha,N}$ is labelled with two 
integers,
$\{n,i\}$, where $0\le n<\verb-INT_MAX-$, $0\le i< N$. 
Any label $\{n,i\}$ corresponds to the membrane voltage $V_{n,i}$ from $\mathbb{V}_{\alpha,N}$:
$
V_{n,i}=\alpha^{n+1} V_0+i\,c_n
$,
or:
\begin{equation}\label{ni2V}
V_{n,i}=\alpha^n V_0
\left(\alpha+\frac{i}{N}(1-\alpha)\right)
\end{equation}
From the last, it is clear that pure decay evolution for time step $dt$
transforms voltage $V_{n,i}$ into $V_{n+1,i}$, which means for the integer labels:
\begin{equation}\label{ni2ni}
\{n,i\}\quad 
\xrightarrow{~dt~}
\quad \{n+1,i\}
\end{equation}

Now, if initial state of neuron is given as floating point number 
$V\in[0;V_0[$, then one needs to find its approximation
 $V_{n,i} \in \mathbb{V}_{\alpha,N}$, 
 as described above. The precision of this approximation is:
\begin{equation}\label{precN}
|V-V_{n,i}|<\Delta  V_0/N
\end{equation}

In the approximation of $V$ with $V_{n,i}$, 
the value of $n$ should satisfy the following relation:
$
\alpha^{n+1}V_0\le V < \alpha^n V_0
$,
which gives:
\begin{equation}\label{V2n}
n = -\left[
\log_\alpha(V_0/V)\right]-1
\end{equation}
where $[x]$ denotes the integer part of $x$.
Now, with $n$ found, we determine the $i$ value from the following 
relation:
$
\alpha^{n+1}V_0+ic_n\le V<\alpha^{n+1}V_0+(i+1)c_n
$,
which gives:
\begin{equation}\label{Vn2i}
i=\left[\left(V-\alpha^{n+1}V_0\right)/c_n\right]
\end{equation}

Now evolution of LIF state can be expressed in integer numbers as follows.
For initial value of voltage, $V(0)<V_0$, we find its integer 
representation/approximation $\{n,i\}$
in accordance with Eqs. (\ref{V2n}), (\ref{Vn2i}).
The pure decay evolution then goes as displayed in (\ref{ni2ni}).
In order to describe influence of input impulse with magnitude $h$
on the state $\{n,i\}$ we calculate the voltage $V_{n,i}$ in accordance to
(\ref{ni2V}). The voltage after receiving input impulse becomes $V_{new}=V_{n,i}+h$. If $V_{new}\ge V_0$, then neuron 
produces an output impulse and ends in the state ``empty'' with $V=0$. 
Otherwise, the new integer state, $\{n',i'\}$, 
can be found with (\ref{V2n}), (\ref{Vn2i}) applied to the $V_{new}$
instead of $V$.

The procedures given above define the dynamical system intLIF
in which a neuronal state is described with two integers,
$\{n,i\}$, with additional unique state ``empty'' attained just
after firing.

\section{Results}

\subsection{Coding}
Neuronal state is described by three integers:
\verb-int n,i; char empty;- If
\verb-empty == 1- then the neuron is in its resting state
with $V=0$. If \verb-empty == 0- then $V$ can be calculated 
from the \verb-n,i- in accordance with (\ref{ni2V}).
Equation (\ref{ni2V}) gives the following C-code for calculating
$V$ from known integer state \verb-{n,i,empty}-:
\begin{verbatim}
V = empty ? 0 : pow(al, n)*V0*(al + (double)i/N*(1 - al));
\end{verbatim} 
where \verb-V0- stands for $V_0$ and \verb-al- stands for $\alpha$.
After firing, \verb-empty == 1-.

Equation (\ref{V2n}) gives the following C-code for calculating $n$:
\begin{verbatim}
n = - floor(log(V0/V)/log(al)) - 1;
\end{verbatim}
Similarly for the equations (\ref{cn}), (\ref{Vn2i}).

\subsection{Testing}

For testing the intLIF simulation paradigm, both intLIF and fpLIF models
were stimulated with the same random stream of impulses. The stream
with exponential distribution of inter-spike intervals (ISI) was
generated with random number generator from the GNU Scientific
Library\footnote{See {\tt http://www.gnu.org/software/gsl/}}. 
Generators of three types were used, each with a number of different seeds, 
see Table \ref{T1}.

\begin{table}[h]
\begin{center}
\caption{\label{T1} Parameters of simulating algorithm}
\begin{tabular}{ll}
random number generators & MT19937, knuthran2002, taus113 \\
number of different seeds for each & 10\\
testing duration, real time & 1 hour\\
max value for $N$ & \verb-Nmax- = $10^9$\\
min value for $dt$ & \verb-dtmin- = 0.001 ms\\
initial $N$ & 10\\
initial $dt$ & 0.1, 0.01, 0.001 ms\\
\end{tabular}
\end{center}
\end{table}

Each ISI obtained from the generator as floating point number
was approximated with a value from $\mathbb{T}$
by rounding to the nearest integer number of time steps $dt$:
\begin{verbatim}
double ISI = gsl_ran_exponential (r, mu);
int iISI = rint(ISI/dt);
ISI = dt*iISI;
\end{verbatim}
where \verb-gsl_ran_exponential (r, mu)- --- is the generator,
\verb-r- --- is the pointer to global generator,
\verb-mu- --- is the mean inter-spike interval of the exponential 
distribution; \verb-mu- $=1/\lambda$, where $\lambda$ is the 
intensity of Poisson stream, which is obtained from the generator
and used as input stimulating stream.
This way obtained ISIs were used to apply input impulses
to both the intLIF and fpLIF model within a single program. 
The program starts with initial values for $N$, $dt$ given in the Table \ref{T1}.
With each input impulse, it is checked whether both fpLIF and intLIF
react the same way, namely, either both fire, or both not fire.
If the reaction for some input impulse is not the same, a new value for $N$
is chosen by multiplying the current value by 10, and simulation starts anew from the 
beginning. If the maximum value for $N$ is
reached (see Table \ref{T1}) and fpLIF and intLIF firing moments 
still do not the same through the whole simulation time, 
next option is to divide $dt$ by 10.
A single run of the program is considered successful if during 1 hour
of real time all input impulses produced the same reaction in both intLIF
and fpLIF.

Several sets of physical parameters, which cover a physiologically 
reasonable range, were used in the testing, see Table \ref{T2}.

\begin{table}[h]
\begin{center}
\caption{\label{T2} Physical parameters used in simulation.}
\begin{tabular}{ll}
threshold depolarization, $V_0$ & 20 mV \\
input impulse height, $h$ & 0.25, 0.5, 1., 2., 4., 8., 16. mV\\
membrane relaxation time, $\tau$ & 10., 20., 40. ms\\
input stream intensity, $\lambda$ & 0.4, 0.8, 1.60, 3.20, 6.40 ms$^{-1}$\\
\end{tabular}
\end{center}
\end{table}

The testing was made for any combination of parameters from both 
Table 1 and 2. Some characteristic numbers of input and output
streams are given in Table \ref{T3}.

\begin{table}[h]
\begin{center}
\caption{\label{T3} Characteristics of streams of impulses during 1 hour of real time testing.}
\begin{tabular}{ll}
minimal number of input impulses & $\sim1.4\cdot 10^6$ \\
maximal number of input impulses & $\sim23.4\cdot 10^6$ \\
minimal number of output impulses & 0 \\
maximal number of output impulses & $\sim11.7\cdot 10^6$ \\
\end{tabular}
\end{center}
\end{table}

As a result of testing, 
it was found that for any combination of parameters it is
possible to ensure that all firing moments of intLIF and fpLIF are identical 
by choosing proper $N$ and $dt$ values. The decisive factor, which
determines whether all spiking moments of intLIF and fpLIF coincide,
is the precision of approximating the interval of possible voltages,
$[0;V_0[$, with discrete values from $\mathbb{V}_{\alpha,N}$ as compared to
 $h$. This relative error, $\delta V$, can be estimated based on 
(\ref{error1}) and (\ref{precN}):
$
\delta V = (1-\alpha)V_0/(Nh),
$
which for small $\frac{dt}{\tau}$ gives:
$
\delta V = (dt\, V_0)/(\tau\,N\, h)
$.
In the testing performed, it appeared that having
$
\delta V\le 2.0\cdot 10^{-11}
$
guarantees that sequences of spiking moments of intLIF and fpLIF are
identical. For larger values of $\delta V$, differences between the two sequences may appear, which are characterized by a few misplaced spikes, up to several hundred and more.

\section{Discussion}
In numerical simulation of a dynamical system adaptive algorithms
are normally used, when the time step $dt$ value is increased/decreased 
during calculations in order to make calculations faster
and more precise. This works perfect if it is necessary to calculate
the system's state at some future moment of time starting from some
initial state. To determine periodic regimes in a reverberating network,
it is instead necessary to calculate the whole trajectory during some interval of time.
In this case, the straightforward way is to approximate that
interval with equidistant discrete points as in (\ref{moments})
and calculate states of the system in those points. Therefore, paradigm
of fixed time step is used here for a single neuron.

Description of neuronal state (voltage) 
with a pair of integers $\{n,i\}$ does not exempt the intLIF model
from using machine floating point numbers. Indeed, in Eqs. 
(\ref{ni2V}), (\ref{V2n}), (\ref{Vn2i}), operations with floating point
numbers are explicitly involved. Nevertheless, the pure decay evolution,
as it is described in (\ref{ni2ni}) goes without rounding errors.
The underlying to (\ref{ni2ni}) values from $\mathbb{V}_{\alpha,N}$
are always the same for the same $\{n,i\}$. With obtained input impulse,
calculation of $V_{n,i}+h$ involves a rounding error. The error
is immediately cleared while calculating new $\{n,i\}$ by means of
(\ref{V2n}), (\ref{Vn2i}). This allows describing states of LIF
neuron in terms of integers in a consistent manner. Namely,
different state labels $\{n,i\}$ correspond to different voltages 
from $\mathbb{V}_{\alpha,N}$ and vice versa.

We used here the simplest possible model for LIF neuron. It seems, that
technique offered in nn. \ref{Pdd}, \ref{Ii}, above, can be extended to be
valid for more elaborated LIF models, like those described in
\cite{Burkitt,Burkitta,Lansky1984,Tuckwell1988}.

\section{Conclusions}

\noindent
The intLIF paradigm for numerical simulation of leaky integrate 
and fire neuron is proposed. In this paradigm, neuronal state is
described by two integers, $\{n,i\}$. The membrane voltage of LIF
neuron can be calculated from $\{n,i\}$, if required. The LIF
state change due to both leakage and stimulating impulses is
expressed exclusively in terms of changing integers $\{n,i\}$. 
The intLIF paradigm is compared with the standard fpLIF simulation paradigm,
where membrane voltage is expressed as machine floating point number,
by stimulating both models with the same random stream of input impulses
and registering the spiking moments of both models.
It is concluded that approximation parameters of intLIF can be chosen
in such a way that spiking moments of both models are exactly the same
if expressed as whole number of simulation time step, $dt$.
Description of LIF states by integers gives a consistent numerical 
model, suitable for situations where exact comparison of states is necessary.
\bigskip

Acknowledgment. This work was supported by the Programs of the NAS of Ukraine
``Microscopic and phenomenological models of fundamental physical
processes in a micro and macro-world'', PK No 0112U000056, and
``Formation of structures in quantum and classical equilibrium
and nonequilibrium systems of interacting particles'', PK No 0107U006886.

\end{document}